\documentclass{osa-article}
\usepackage[T1]{fontenc}
\usepackage{physics}
\usepackage{subcaption}
\usepackage{amsmath}
\usepackage{siunitx}
\journal{osajournal}

\articletype{Research Article}

\begin{document}
\UseRawInputEncoding

\title{All-optical input-agnostic polarization transformer via experimental Kraus-map control}

\author{Wenlei Zhang, \authormark{1,+} Ravi K. Saripalli, \authormark{1,+} Jacob M. Leamer, \authormark{1} Ryan T. Glasser, \authormark{1,2} Denys I. Bondar \authormark{1,3}}

\address{\authormark{1}Department of Physics and Engineering Physics, Tulane University, 6823 St. Charles Avenue, New Orleans, 70118, USA\\
\authormark{+}these authors contributed equally to this work\\
\authormark{2}rglasser@tulane.edu\\
\authormark{3}dbondar@tulane.edu}




\begin{abstract*}
The polarization of light is utilized in many  technologies throughout  science and engineering. The ability to transform one state of polarization to another is a key enabling technology. Common polarization transformers are simple polarizers and polarization rotators. Simple polarizers change the intensity depending on the input state and can only output a fixed polarized state, while polarization rotators rotates the input Stokes vector in the 3D Stokes space. We experimentally demonstrate an all-optical input-agnostic polarization transformer (AI-APT), which transforms all input states of polarization to a particular state that can be polarized or partially polarized. The output state of polarization and intensity depends solely on setup parameters, and not on the input state, thereby the AI-APT functions differently from simple polarizers and polarization rotators. The AI-APT is completely passive, and thus can be used as a polarization controller or stabilizer for single photons and ultrafast pulses. To achieve this, we, for the first time, experimentally realize complete kinematic state controllability of an open single-qubit by Kraus maps put forth in [Wu {\it et al.} J. Phys. A {\bf 40}, 5681 (2007)]. The AI-APT may open a new frontier of partially polarized ultrafast optics.
\end{abstract*}

\section{Introduction}

The polarization of light describes the trajectory of the electric field vector as it oscillates. Technologies based on polarization are being used in many fields, such as machine vision \cite{wolff_polarization-based_1990,koshikawa_model-based_1987}, remote sensing \cite{tyo_review_2006,schott_fundamentals_2009}, biomedical optics \cite{ghosh_tissue_2011}, astronomy \cite{vlemmings_2007,deglinnocenti_polarization_2004}, and free-space optical communication \cite{Zhang:17,Ma:17,10.1117/12.2312038}. Many quantum information technologies also utilize the polarization degree-of-freedom of light \cite{gasparoni_realization_2004,kok_linear_2007,obrien_photonic_2009,kagalwala_single-photon_2017}.
In addition, ultrafast vector beams \cite{Ravi_SHGvector_2019} with tailored spatial and temporal polarization shaping \cite{dorney_controlling_2019} have the potential to be used in studies of chiral molecules and magnetic materials \cite{alonso_complete_2020}.
In all of these varied applications of polarization, the ability to transform one state of polarization (SOP, or simply ``state'') to another is a key enabling technology. \par
The most common components/devices used for polarization transformation are simple linear/circular polarizers and polarization rotators. Simple polarizers are usually made of a thin film of polarizing material whose transmittance is polarization-dependent \cite{banning_practical_1947}. As a result, the output intensity of such polarizers depends on the input state. The output state is always polarized, even if the input state is partially polarized. Polarization rotators usually consist of components that change the SOP based on birefringence, on the Faraday effect, or on total internal reflection \cite{duarte_tunable_2015}. Such rotators preserve the input degree of polarization (DOP) and intensity and rotate the normalized input Stokes vector in the 3D Stokes space (the Poincar\'{e} sphere picture). The output state of a polarization rotator depends on both the device settings and the input state. Both types of polarization transformers, simple polarizers and polarization rotators, are commonly used in optical setups involving the polarization degree-of-freedom. A new generation of polarization transformers based on metasurfaces \cite{dorrah_metasurface_2021}, which enables propagation-dependent polarization responses without \textit{a priori} knowledge of the incident state, is currently being developed. Another example of a metasurface polarizer can transform all input states into any state on the Poincar\'{e} sphere with transmittances depending on the input state \cite{wang_arbitrary_2021}. \par
Here, we present optical setups for an all-optical input-agnostic polarization transformer (AI-APT) that has different functionality from the previously mentioned polarization transformers. The AI-APT transforms all input states into a particular fixed target output state, which can be polarized or partially polarized, without polarization-dependent losses. The output intensity and output state depend solely on the settings of the AI-APT setup and not on the input state. As a demonstration, we experimentally measured the output states of the AI-APT for various input states, which confirmed that the output states are independent of the input states. \par
Unlike common polarizers and polarization rotators, the AI-APT is a completely passive device that performs an all-to-one mapping on the polarization state. Thus, it could be used as a polarization controller or stabilizer, especially for polarization qubits encoded in single photons and ultrafast pulses. Most existing polarization stabilizers that reduce noise of the input state rely on measurement and feedback schemes \cite{koch_versatile_2014,chiba_polarization_1999,martinelli_polarization_2006}. However, such schemes do not work for single photons, where the photon is lost upon measurement, or ultrafast pulses, where the pulse duration is significantly shorter than the response time of feedback electronics. There are other proposed types of polarization stabilizers that utilize polarization attraction based on non-linear effects \cite{millot_nonlinear_2014}, but they require a non-linear medium and a strong auxiliary pump beam of a different wavelength, which greatly increases the complexity in implementation. Another type of nonlinear omnipolarizer based on optical fibers has also been proposed \cite{fatome_universal_2012}, which can transform any input state into either left- or right-circular polarization. The AI-APT we propose could enable access to new areas of research involving ultrafast optical pulses with incoherent polarization. For example, the decoherence of ultrashort extreme-ultraviolet pulses could be used for super-resolution photoelectron spectroscopy and jitter-free experiments \cite{bourassin-bouchet_quantifying_2020}. Incoherent polarization can also be used for studies of spin-related phenomena and optical manipulation \cite{eismann_transverse_2021}. \par
Our AI-APT setup shown in this work is inspired by the concept of the Kraus-controllability and the mathematical equivalence between the density matrix of a single qubit and the coherency matrix of classical light. To the best of our knowledge, our experimental results are also the first experimental demonstration of the Kraus-controllability for a open single-qubit. For finite-dimensional open quantum systems, whose dynamics are described by Kraus maps \cite{kraus_states_1983}, it has been proven that there exists a Kraus map which transforms all initial states (pure or mixed) into a predefined target state (pure or mixed) \cite{wu_controllability_2007}. Compared with unitary control, such Kraus-controllability is robust against variations in the initial state. The Kraus-controllability may be utilized in many quantum information protocols. For example, it may be used in initial state preparation to reduce the effect of noise. It may also be used in quantum information dilution \cite{ziman_diluting_2002,roa_measurement-driven_2006} to realize a mapping of an unknown quantum state to a given target state. 

\section{The coherency matrix}

The SOP can be described by the Stokes vector \cite{stokes_2009,born_principles_1999} or the coherency matrix \cite{goodman_statistical_2000}, which is a generalization of the Jones calculus \cite{jones_new_1941}. The Stokes vector, $(s_0,s_1,s_2,s_3)$, and the coherency matrix, $\rho$, are related by

\begin{gather}
    \rho=\frac{1}{2}\begin{pmatrix}
    s_0+s_1 & s_2+is_3 \\
    s_2-is_3 & s_0-s_1
    \end{pmatrix}. \label{eq:jones}
\end{gather}
The coherency matrix provides all second-order statistical information about the polarization state. The DOP of $\rho$ is given by

\begin{gather}
    P=\frac{\sqrt{s_1^2+s_2^2+s_3^2}}{s_0}.
\end{gather}
\par
The equivalency between the coherency matrix and the density matrix of a quantum two-level system has been well-established \cite{falkoff_stokes_1951,fano_stokes-parameter_1954,parrent_matrix_2008}. It has also been proven that there exist Kraus-maps from all density matrices to a particular density matrix, which can be either pure or mixed \cite{wu_controllability_2007}. We applied these results to the coherency matrix to develop the AI-APT setups. Let us summarize the relevant steps. In the following text, we will use $\ket{0}$ and $\ket{1}$ to respectively denote the Jones vectors of horizontal and vertical linear polarizations. \par
For any pure state $\rho_{p}=\ket{\psi}\bra{\psi}$, where $\ket{\psi}=U_{0}\ket{0}$ and $U_{0}$ is an $\mathrm{SU}(2)$ matrix, we can construct a Kraus-map $\Phi(\rho)=\sum_{i}K_{i}{\rho}K_{i}^{\dagger}$ such that $\Phi(\rho)=\rho_{p}$ for all $\rho$ \cite{wu_controllability_2007}. One simple choice is 
\begin{gather}
    K_{1}=\ket{\psi}\bra{0}, \\
    K_{2}=\ket{\psi}\bra{1}, \\
    \Phi(\rho)=\ket{\psi}\bra{0}\rho\ket{0}\bra{\psi}+\ket{\psi}\bra{1}\rho\ket{1}\bra{\psi}=\Tr(\rho)\ket{\psi}\bra{\psi}=\rho_{p}.
\end{gather}
It is easy to check that the operation elements satisfy the Kraus-map condition $\sum_{i=1}^{2}K_{i}^{\dagger}K_{i}=I$, where $I$ is the identity matrix. We can rewrite $K_1$ and $K_2$ in the following form so that it is easy to see what optical elements are required to implement these operations,
\begin{gather}
    K_{1}=U_{0}\ket{0}\bra{0}, \label{eq:K1} \\
    K_{2}=U_{0}\ket{0}\bra{1}=U_{0}\sigma_{1}\ket{1}\bra{1}, \label{eq:K2}
\end{gather}
where $\sigma_{1}$ is the first Pauli matrix.

\section{Experimental setups and results}

We will present the AI-APT setup in separate cases where the target output state is polarized or partially polarized. We will also show the experimentally measured output states for various input states in each case. 

\subsection{Polarized target state}

\begin{figure}[htbp]
    \centering
    \includegraphics[width=\linewidth]{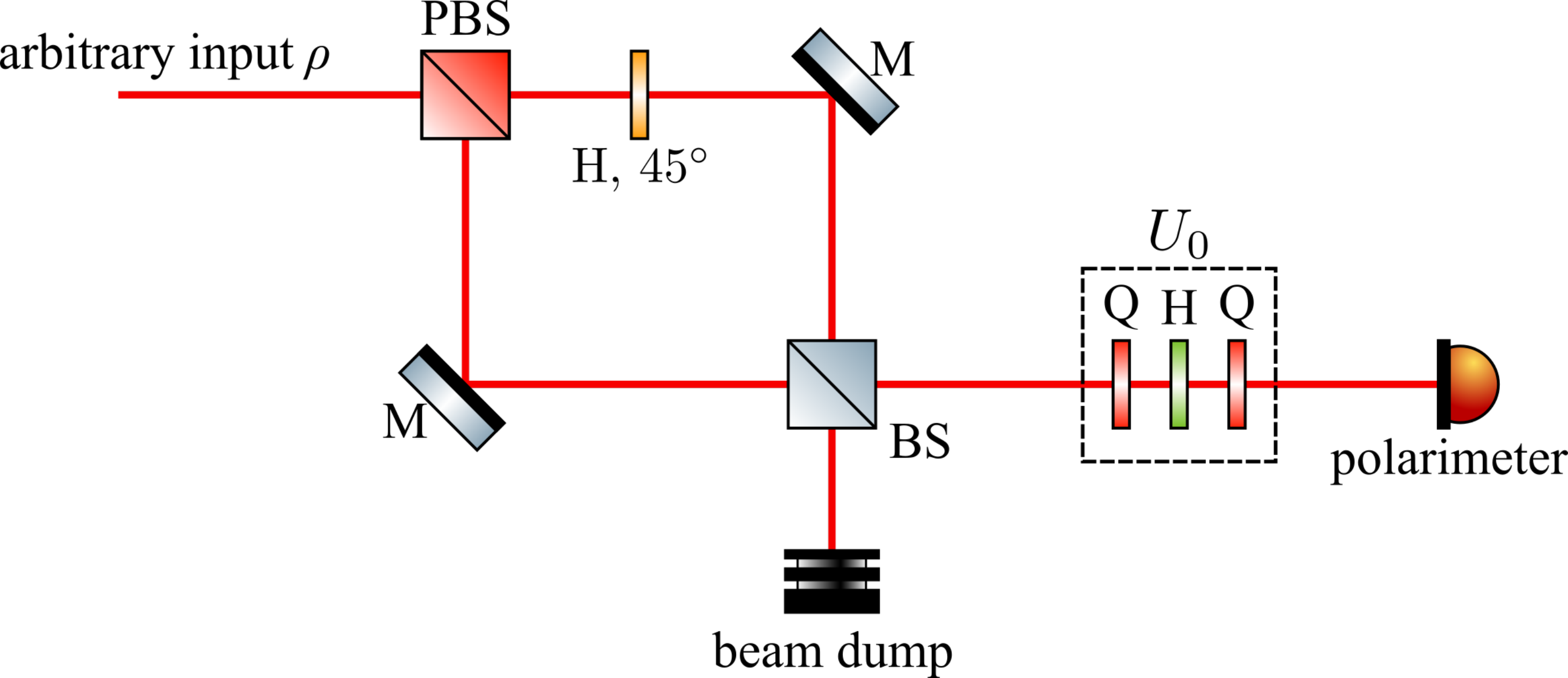}
    \caption{Experimental setup of the AI-APT for a polarized target state. PBS: polarizing beamsplitter; BS: 50/50 non-polarizing beamsplitter; Q: quarter-wave plate; H: half-wave plate; M: mirror.}
    \label{fig:pure_setup}
\end{figure}

Figure \ref{fig:pure_setup} shows the setup of the AI-APT for a polarized target state, which is obtained from Eqs.~\eqref{eq:K1} and \eqref{eq:K2}. The input state goes through a Mach-Zehnder interferometer (MZI) with a polarizing beamsplitter (PBS) and a 50/50 non-polarizing beamsplitter (BS). The transmitted light from the PBS is horizontally-polarized, while the reflected light is vertically-polarized. In the transmitted path, a half-wave plate (HWP) with its fast-axis at $45^\circ$ w.r.t. the horizontal axis realizes the transformation corresponding to $\sigma_{1}$. Since both states into the BS are always vertically-polarized, the state after the BS is also always vertically-polarized. An $\mathrm{SU}(2)$ rotation $U_0$ in the Stokes space, implemented by three waveplates \cite{simon_universal_1989,simon_minimal_1990}, transforms the vertical polarization to the desired state. The state detected by the polarimeter only depends on $U_0$, and not on the input state. The only loss incurred, apart from experimental losses such as reflective losses from linear optical elements, is from the undetected port of the BS, thus the relative output intensity, given by $s_0'/s_0$, where $s_0'$ is the intensity measured by the polarimeter and $s_0$ is the input intensity, is always equal to $1/2$, regardless of the input state. \par

\begin{figure}[htbp]
    \centering
    \includegraphics[width=\linewidth]{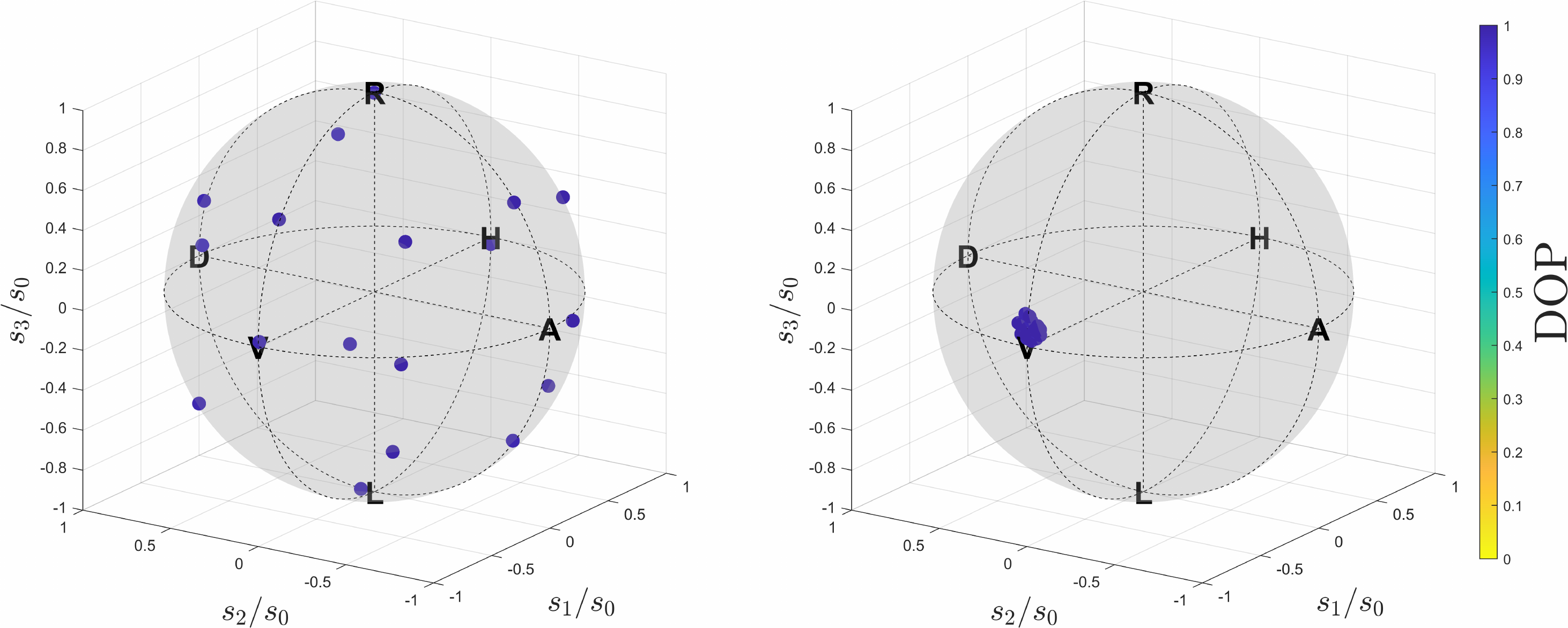}
    \caption{Measured input (left) and output (right) states of the pure-state AI-APT. The $\mathrm{SU}(2)$ rotation $U_0$ is chosen to be the identity matrix. The mean DOP of all output states is $0.996$.}
    \label{fig:apt_pure_results}
\end{figure}

Figure \ref{fig:apt_pure_results} shows the measured input states and corresponding output states of the setup shown in Fig.~\ref{fig:pure_setup}. The $\mathrm{SU}(2)$ rotation $U_0$ is chosen to be the identity matrix $I$ for simplicity. These results show the AI-APT can transform various input states to the same target state, which cannot be achieved by any $\mathrm{SU}(2)$ rotation. The mean relative intensity is $0.465$
, which is slightly lower than the expected value of 1/2 due to experimental losses. The mean input power is \SI{1.58}{mW}. Apart from reflective losses, another source of loss comes from the interference at BS used to combine the two beams. The two beams into the BS will interfere due to having the same SOP. If slightly misaligned, the two beams will create bright interference fringes, which are hard to collect into the polarimeter. We reduce this loss by exactly overlapping the two beams on the BS, which creates an interference pattern with a bright central lobe and weak and far away outer rings. In this case, almost all optical power will be located in the central lobe, which can be directed into the polarimeter. However, this interference pattern is sensitive to external noises such as vibrations. \par

\subsection{Partially polarized target state}

\begin{figure}[htbp]
    \centering
    \includegraphics[width=\linewidth]{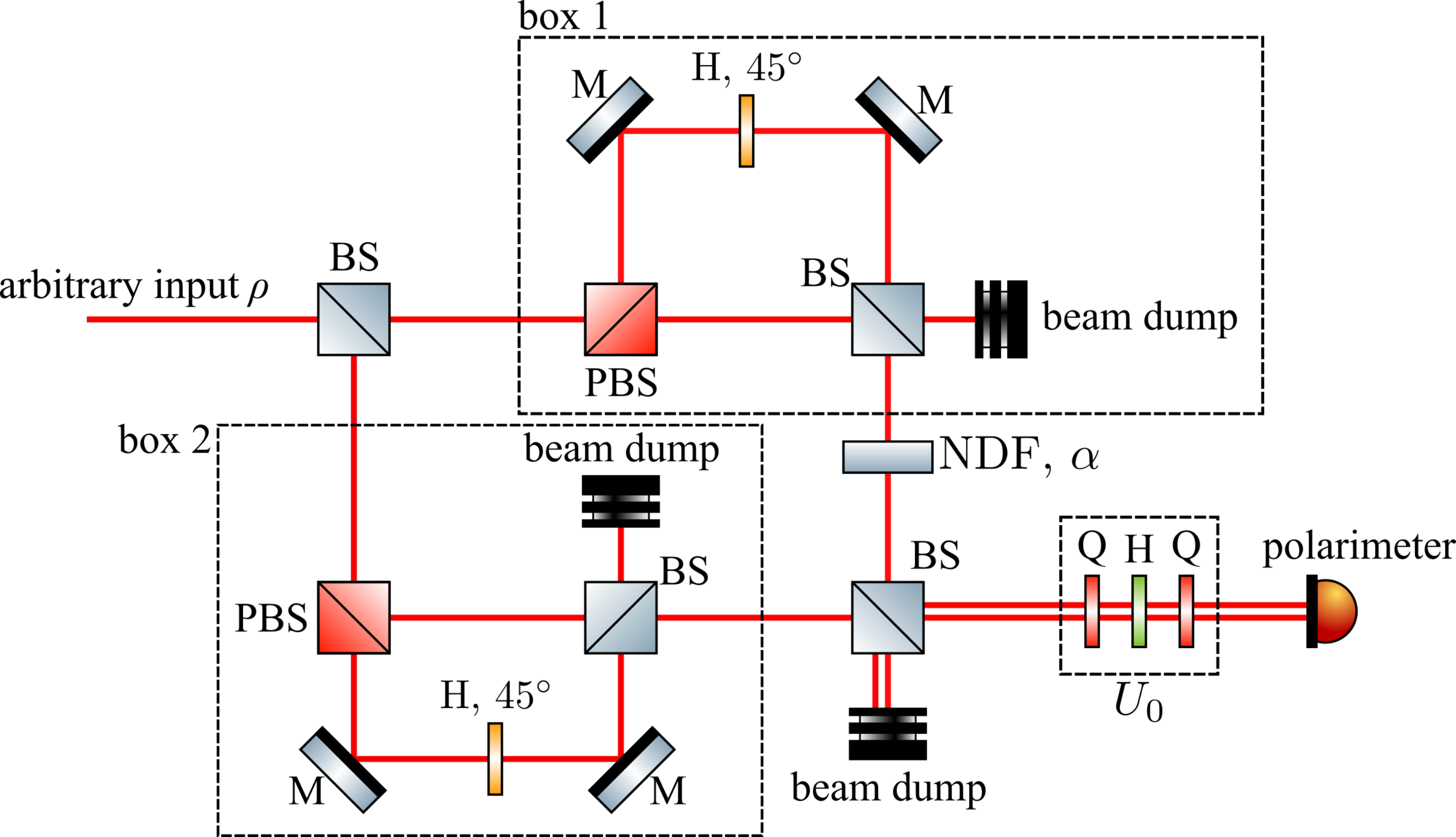}
    \caption{Experimental setup of the AI-APT when the target state is partially polarized. PBS: polarizing beamsplitter; BS: 50/50 non-polarizing beamsplitter; Q: quarter-wave plate; H: half-wave plate; M: mirror; NDF: neutral-density filter.}
    \label{fig:apt_mixed_setup}
\end{figure}

Once we have the setup for an AI-APT with a polarized target state, it is fairly straightforward to combine two AI-APTs with orthogonal polarized target states to form an AI-APT with a partially polarized target state. The setup in Fig.~\ref{fig:pure_setup} can be modified to create an AI-APT with a target state of horizontal polarization by simply moving the HWP to the other arm of the MZI and keeping $U_{0}=I$. Two AI-APTs with respective target states of horizontal polarization and vertical polarization can be combined to form an AI-APT where the target state is any partially polarized state. Such a setup is shown in Fig.~\ref{fig:apt_mixed_setup}. The input state is split into two copies, one of which is transformed into horizontal polarization by the AI-APT in box 1 and the other is transformed into vertical polarization by the AI-APT in box 2. The horizontal polarization state passes through a neutral-density filter (NDF) with transmittance $\alpha$ and combines with the vertical polarization state on the final BS. The two beams into the final BS are aligned such that the two output beams are parallel with a small separation between them. Since the polarimeter does not distinguish between spatial modes, the detected SOP is the incoherent sum of the states of the two separated beams \cite{aiello_linear_2007,barberena_experimental_2015,lizana_arbitrary_2015}. This incoherent sum results in a Stokes vector pointing towards the vertical SOP in Stokes space, but with a DOP that depends on $\alpha$. Finally, the SOP can be rotated to the desired direction by an $\mathrm{SU}(2)$ rotation $U_{0}$. The output state detected by the polarimeter is solely determined by the setup parameters $\alpha$ and $U_0$. The losses in this setup, apart from the experimental losses mentioned previously, come from the undetected ports of the beamsplitters and the absorption and/or reflection of the NDF, which are both independent of the input state. Note that the NDF can be moved to the other input of the final BS, which will lead to a Stokes vector in the opposite direction, and the change in direction can be compensated by $U_{0}$. The state before $U_0$ can be described by the following expression \cite{barberena_experimental_2015},
\begin{gather}
    \ket{\Psi}=\sqrt{\frac{\alpha}{\alpha+1}}\ket{H}\ket{X}_\mathrm{aux}+\sqrt{\frac{1}{\alpha+1}}\ket{V}\ket{Y}_\mathrm{aux},
\end{gather}
where $\ket{H}$ and $\ket{V}$ represent the horizontal and vertical polarization states, respectively, and $\ket{X}_\mathrm{aux}$ and $\ket{Y}_\mathrm{aux}$ represent the paths of the two spatially-separated beams acting as an auxiliary degree of freedom. Using the purification theorem \cite{kirkpatrick_schrodinger-hjw_2006}, the SOP obtained by tracing over the auxiliary degree of freedom (i.e., $\rho=\Tr_\mathrm{aux}\ket{\Psi}\bra{\Psi}$) will be a partially polarized state. The required tracing over the auxiliary degree of freedom is realized by measuring the SOP with a polarimeter that does not distinguish between spatial modes. \par
In the setup shown in Fig.~\ref{fig:apt_mixed_setup}, for an input state with the Stokes vector $(s_0,s_1,s_2,s_3)$, the Stokes vector of the output of box 2 is $(1/4)(s_0,-s_0,0,0)$ and the Stokes vector of the output of box 1, after the NDF, is $(1/4)(\alpha{s_0},\alpha{s_0},0,0)$. The final BS sums these two Stokes vector to obtain $(1/8)((\alpha+1)s_0,(\alpha-1)s_0,0,0)$, the DOP and relative intensity of which are given by, respectively,
\begin{gather}
    P=\frac{1-\alpha}{\alpha+1}, \label{eq:apt_mixed_DOP} \\
    \frac{s_0'}{s_0}=\frac{1}{8}(\alpha+1). \label{eq:intensity}
\end{gather} \par

\begin{figure}[htbp]
    \centering
    \includegraphics[width=\linewidth]{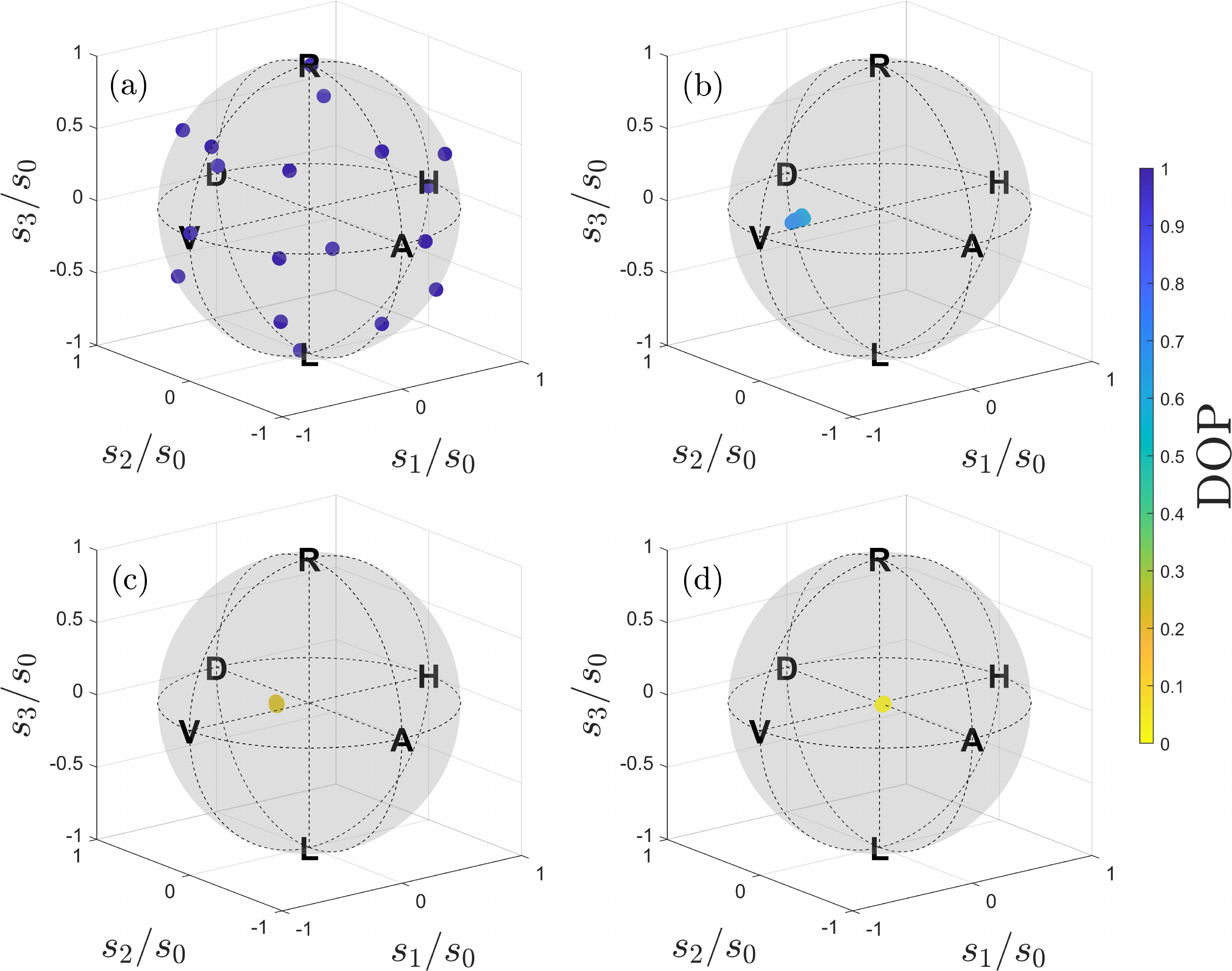}
    \caption{The measured (a) input states of the partially-polarized AI-APT and the corresponding output states for (b) $\alpha=0.22$, (c) $\alpha=0.62$, and (d) $\alpha=0.96$. The $\mathrm{SU}(2)$ rotation $U_0$ is chosen to be the identity matrix. The color of each point represents the DOP of the respective state.}
    \label{fig:apt_mixed_results}
\end{figure}

\begin{figure}[htbp]
    \centering
    \begin{subfigure}{0.48\linewidth}
        \centering
        \includegraphics[width=\linewidth]{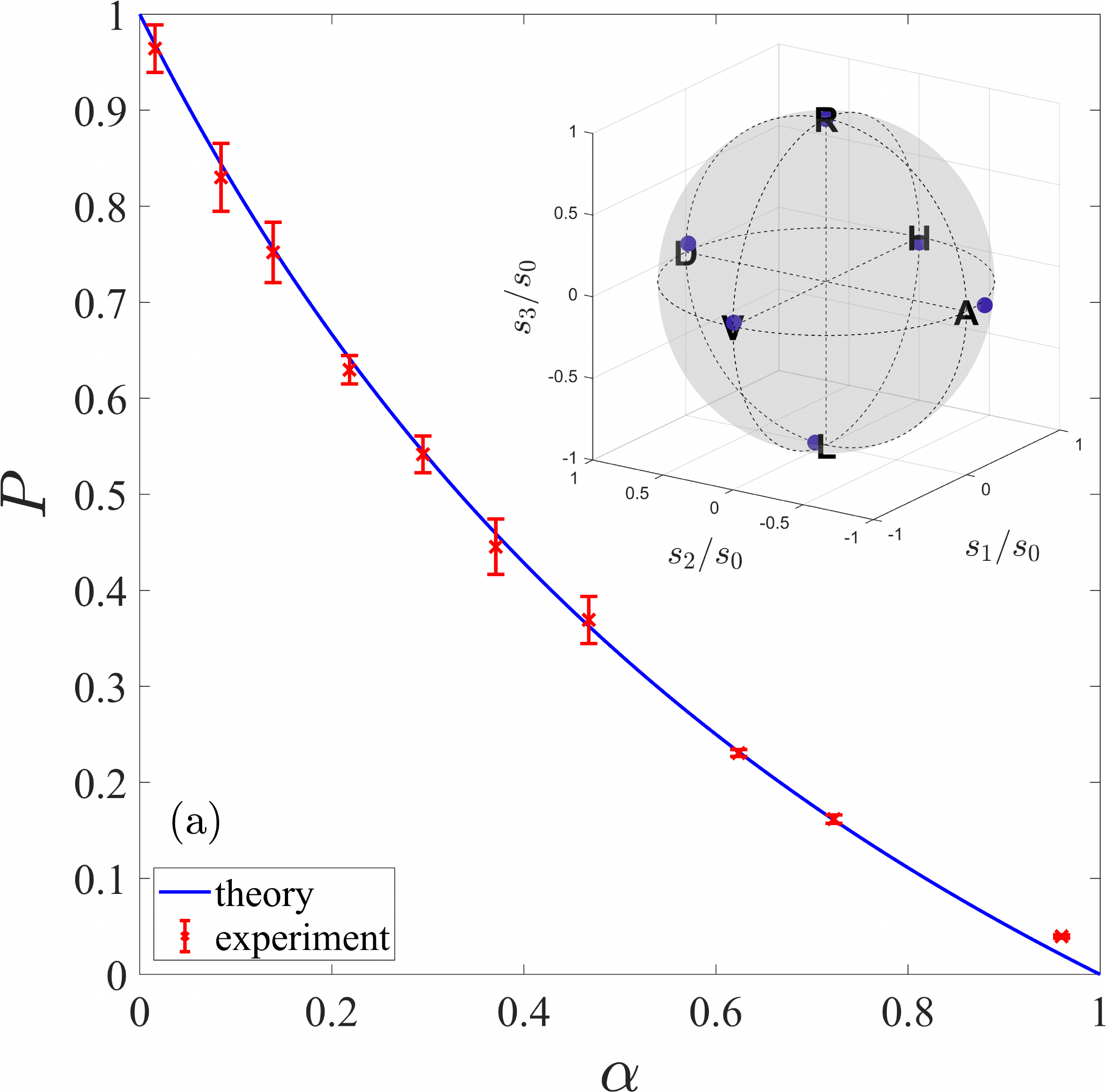}
    \end{subfigure}
    \begin{subfigure}{0.48\linewidth}
        \centering
        \includegraphics[width=\linewidth]{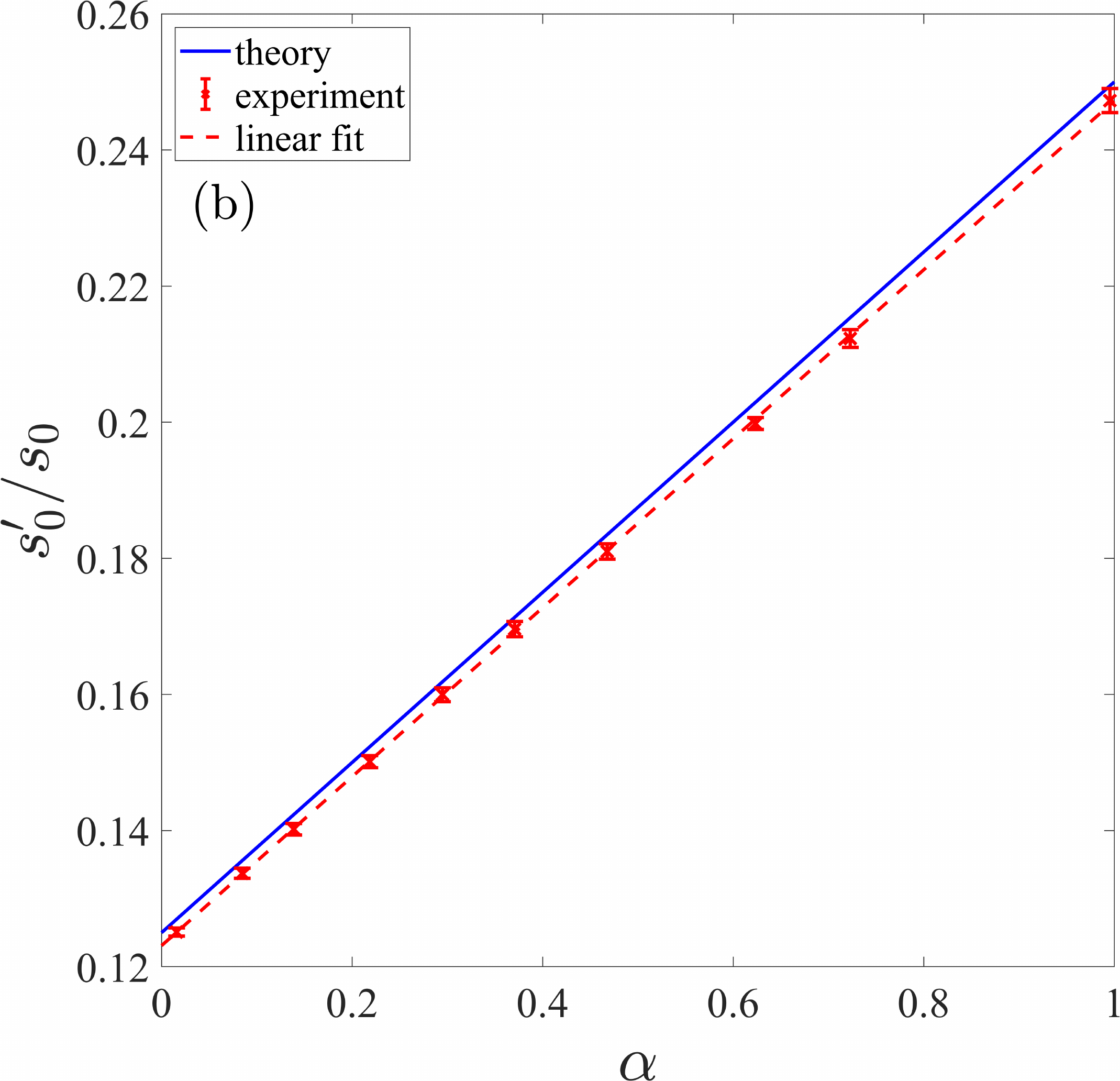}
    \end{subfigure}
    \caption{The mean values of (a) DOP and (b) relative intensity of output states of the partially-polarized AI-APT for various values of $\alpha$. The measurements were made with 6 different input states, which are shown in the inset of (a). The theory curves in (a) and (b) are calculated from Eq.~\eqref{eq:apt_mixed_DOP} and Eq.~\eqref{eq:intensity}, respectively. The dashed line in (b) is the linear fit of the experimental measurements. The error bars represent 95\% confidence interval.}
    \label{fig:apt_mixed_DOP}
\end{figure}

Figure \ref{fig:apt_mixed_results} shows the measured input states and corresponding output states of the AI-APT with partially-polarized target state for various values of $\alpha$ and $U_{0}=I$. These results show the AI-APT is able to transform different polarized input states on the Poincar\'{e} sphere to nearly the same partially-polarized state inside the Poincar\'{e} sphere with a DOP depending on the value of $\alpha$. We also measured the output DOPs and relative intensities for more values of $\alpha$ and compared them to the theoretically expected values. These results are shown in Fig.~\ref{fig:apt_mixed_DOP}. In Fig.~\ref{fig:apt_mixed_DOP}, the experimental results closely match the expected values. For all values of $\alpha$, the relative intensities are lower than the expected values due to experimental losses not taken into account in the theoretical calculation. For the measurements shown in Fig.~\ref{fig:apt_mixed_results} and Fig.~\ref{fig:apt_mixed_DOP}, the input power is kept at \SI{2.61}{mW}.  The main source of experimental losses are the same as those discussed in the setup for the polarized target state (Fig.~\ref{fig:pure_setup}). 

\begin{figure}[htbp]
    \centering
    \includegraphics[width=\linewidth]{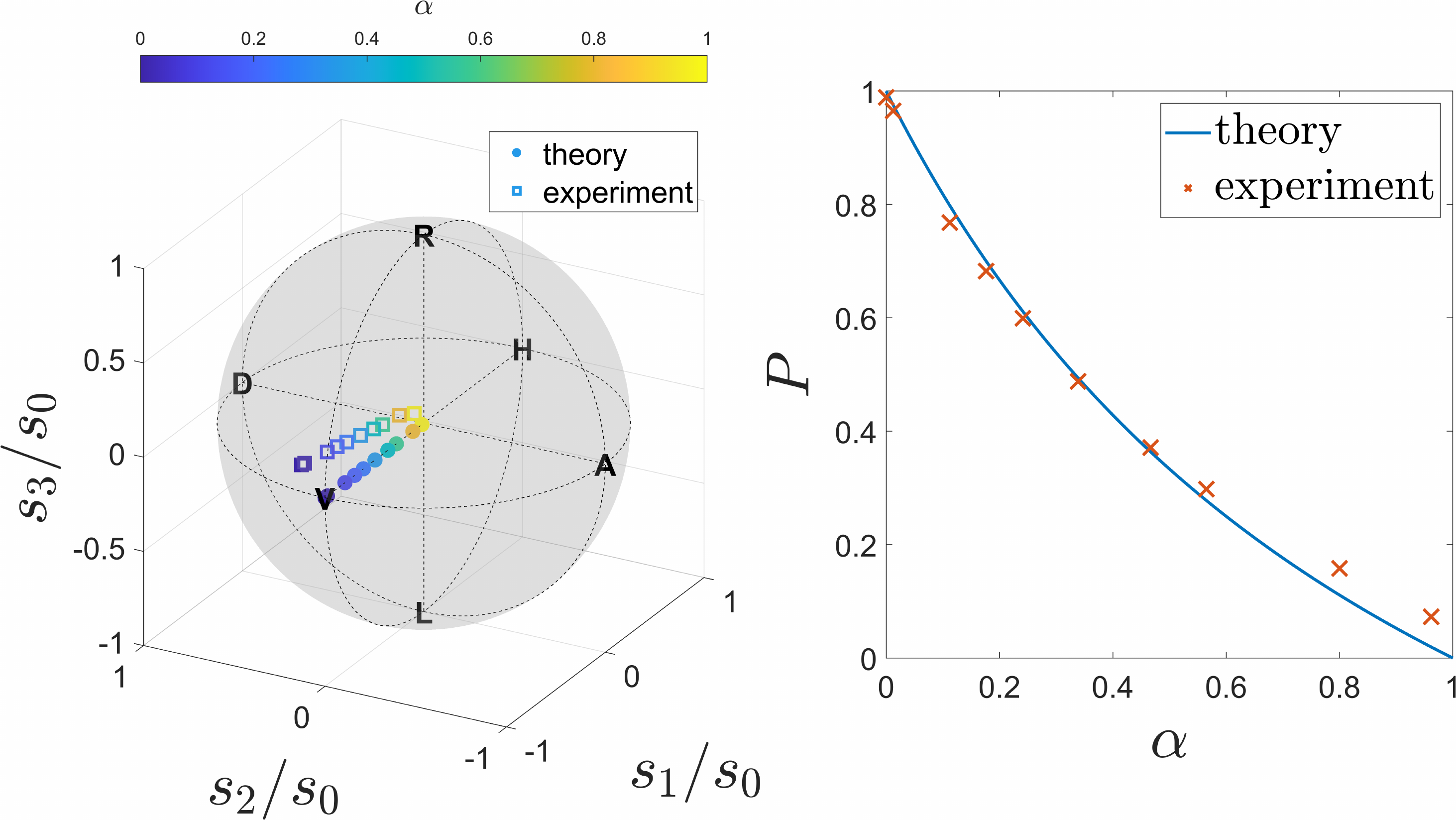}
    \caption{Measured output states of the mixed-state AI-APT for a partially-polarized input state $\rho_{1}$ and $U_{0}=I$ for various values of $\alpha$. Left: the measured output states in Stokes space; right: the output DOPs for each $\alpha$ value. The theory curve is calculated from Eq.~\eqref{eq:apt_mixed_DOP}.}
    \label{fig:unpolarized_input}
\end{figure}

We have also measured the partially-polarized AI-APT outputs for a partially-polarized input state $\rho_{1}$ and $U_{0}=I$, the results of which are shown in Fig.~\ref{fig:unpolarized_input}. The normalized Stokes vector of $\rho_{1}$ is $(1,0.0044,0.28,0.013)$, and the DOP is 0.28. The measured outputs are close to the expected states. \par

\begin{figure}[htbp]
    \centering
    \includegraphics[width=\linewidth]{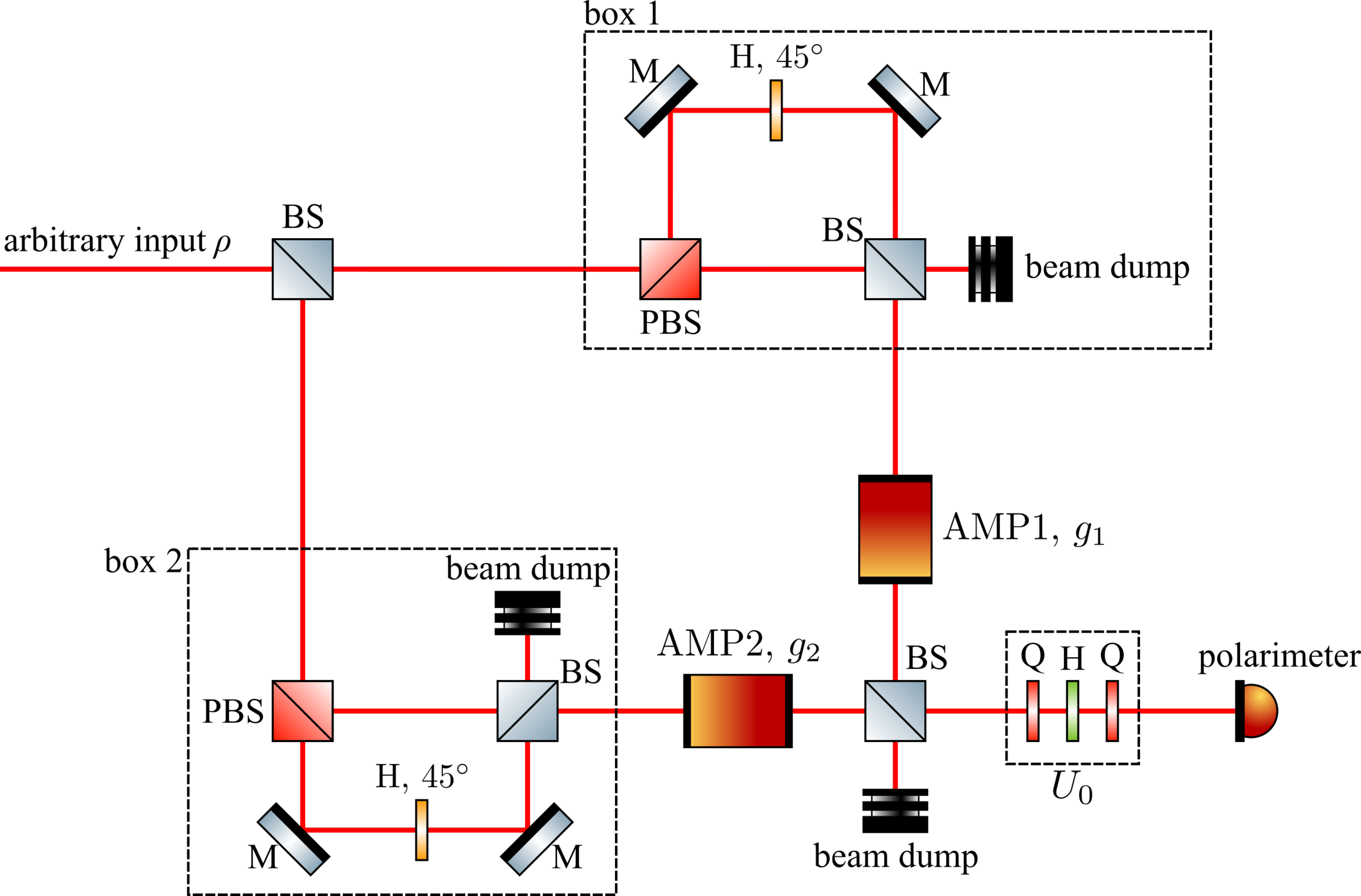}
    \caption{Experimental setup of the AI-APT when the target state is partially polarized with amplifiers to compensate for the loss. PBS: polarizing beamsplitter; BS: 50/50 non-polarizing beamsplitter; Q: quarter-wave plate; H: half-wave plate; M: mirror; AMP: amplifier.}
    \label{fig:mixed_setup_amp}
\end{figure}

Since the SOPs of the beams out of box 1 and box 2 in Fig.~\ref{fig:apt_mixed_setup} are known and fixed regardless of the input state, we can place optical amplifiers at these locations to compensate for the losses incurred throughout the AI-APT setup. Such a setup is shown in Fig.~\ref{fig:mixed_setup_amp}. By choosing appropriate values for the amplifier gains $g_1$ and $g_2$, we can obtain arbitrary output intensity and DOP.

\section{Conclusion}

We presented AI-APT setups in separate cases where the target states are polarized and partially polarized. By experimentally measuring the output SOPs, we showed that we have achieved an all-to-one polarization transformation, i.e. any input polarized state is transformed to a polarized target state or a partially polarized target state. The AI-APT setups presented here can be easily modified to be used for polarization qubits encoded in single photons \cite{aiello_linear_2007}. The AI-APT can enable accurate polarization control for single photons and ultrafast pulses, especially for partially polarized states, and allow us to further explore the DOP degree-of-freedom for various  applications in optical communication \cite{Kikuchi_Ananlysis_2001} and biomedical imaging \cite{Louie_Degree_2018}. When combined with other techniques, such as generating fractional-order polarization singularities \cite{pisanty_knotting_2019} or time-varying orbital angular momentum \cite{rego_generation_2019}, the AI-APT may be used to create various novel complex states of light. Such states could open the way to new directions of research in ultrafast optics.

\begin{backmatter}

\bmsection{Acknowledgments}
This work was supported by the Defense Advanced Research Projects Agency (DARPA) grant number D19AP00043 under mentorship of Dr. Joseph Altepeter. J.M.L. is supported by the Louisiana Board of Regents' Graduate Fellowship Program. R.T.G. also acknowledges funding from the U.S. Office of Naval Research (ONR) under grant N000141912374. D.I.B. is also supported by the U.S. Army Research Office (ARO) under grant W911NF-19-1-0377. The views and conclusions contained in this document are those of the authors and should not be interpreted as representing the official policies, either expressed or implied, of DARPA, ONR, ARO, or the U.S. Government. The U.S. Government is authorized to reproduce and distribute reprints for Government purposes notwithstanding any copyright notation herein.

\bmsection{Disclosures}
The authors declare no conflicts of interest.

\bmsection{Data availability} Data underlying the results presented in this paper are not publicly available at this time but may be obtained from the authors upon reasonable request.

\end{backmatter}



\bibliography{refs}

\end{document}